# Polymeric ruthenium precursor as a photoactivated antimicrobial agent

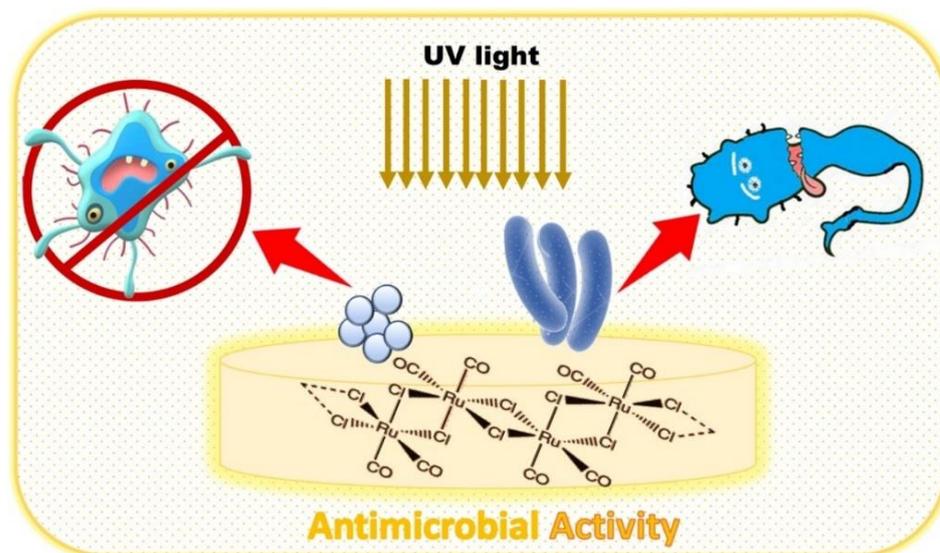

# Polymeric ruthenium precursor as a photoactivated antimicrobial agent

Srabanti Ghosh[1,*], Georgiana Amariei[2], Marta E. G. Mosquera[1,*], Roberto Rosal[2]

[1] Department of Organic and Inorganic Chemistry, Instituto de Investigación en Química "Andrés M. del Río" (IQAR), Universidad de Alcalá, Campus Universitario, 28805, Alcalá de Henares, Madrid-Spain
[2] Department of Chemical Engineering, Universidad de Alcalá, Campus Universitario, 28805, Alcalá de Henares, Madrid-Spain

* Corresponding authors: Dr. Marta E. G. Mosquera, martaeg.mosquera@uah.es & Dr. Srabanti Ghosh, srabanti.ghosh@uah.es

**Abstract**

Ruthenium coordination compounds have demonstrated a promising anticancer and antibacterial activity, but their poor water solubility and low stability under physiological conditions may limit their therapeutic applications. Physical encapsulation or covalent conjugation with polymers may overcome these drawbacks, but generally involve multistep reactions and purification processes. In this work, the antibacterial activity of the polymeric precursor dicarbonyldichlororuthenium (II) $[Ru(CO)_2Cl_2]_n$ has been studied against *Escherichia coli* and *Staphylococcus aureus*. This Ru-carbonyl precursor shows minimum inhibitory concentration at nanogram per millilitre, which renders it a novel antimicrobial polymer without any organic ligands. Besides, $[Ru(CO)_2Cl_2]_n$ antimicrobial activity is markedly boosted under photoirradiation, which can be ascribed to the enhanced generation of reactive oxygen species under UV irradiation. $[Ru(CO)_2Cl_2]_n$ has been able to inhibit bacterial growth via the disruption of bacterial membranes and triggering upregulation of stress responses as shown in microscopic measurements. The activity of polymeric ruthenium as an antibacterial material is significant even at 6.6 ng/mL while remaining biocompatible to the mammalian cells at much higher concentrations. This study proves that this simple precursor, $[Ru(CO)_2Cl_2]_n$, can be used as an antimicrobial compound with high activity and a low toxicity profile in the context of need for new antimicrobial agents to fight bacterial infections.

## 1. Introduction

Antimicrobial resistance is at the foremost of global concerns since common bacterial pathogens like *Staphylococcus aureus*, *Streptococcus pneumoniae* and *Escherichia coli*, have become progressively more to traditional antibiotics causing serious morbidity and mortality (Burnham et al., 2017; Coates et al., 2002; Furuya and Lowy, 2006; Levy and Marshall, 2004). Therefore, for the past decades, considerable research has been done on the design and development of new antimicrobials (Rasko and Sperandio, 2010; Ghosh et al., 2016; Amariei et al., 2018; Ghosh et al., 2011; Kaviyarasu et al., 2017a; Kaviyarasu et al., 2017b; Magdalane et al., 2018). Metal-based drugs have demonstrated outstanding biochemical activities as well as therapeutic actions, and transition metal complexes have been widely applied in clinical research including as anti-cancer, anti-bacterial drugs etc. (Nardon and Fregona, 2018; Bharti and Singh, 2009; Gianferrara et al., 2009).

In particular, ruthenium and ruthenium complexes have demonstrated significant antimicrobial activities and low cytotoxicity (Li et al., 2015; Southam et al., 2017). Besides, some ruthenium complexes have specific optical, electrochemical properties with a strong absorption band in the visible light range due to metal-to-ligand charge transfer (MLCT) (Ma et al., 2013). Furthermore, ruthenium, a group 8 metal, displays iron mimicking properties such as the interchange between oxidation states II and III when binding to biomolecules and the capability to strongly link to nucleic acids and proteins, while its ligand exchange kinetics is comparable to platinum complexes (Luedtke et al., 2003; Metcalfe and Thomas, 2003; Gill and Thomas, 2012; Ramos et al., 2012).

So far, some ruthenium compounds have been described in antimicrobial studies. The multimodal action of ruthenium complexes has been well recognized in (i) a functional and structural role, where the Ru metal octahedral centre may bind to the biological target *via* non-covalent interactions and subsequently labile ligands dissociate, (ii) the metal centre acting as a carrier for active ligands to boost their pharmaceutical activities, (iii) a metal complex behaving as a catalyst for the glutathione oxidation to glutathione disulphide to give an increase of reactive oxygen species (ROS), and (iv) a metal complex that can be used as a photo-sensitizer to generate singlet oxygen due of their low-energy triplet excited state (Yang et al., 2018). Although, Ru(II) complexes employed as antimicrobials have demonstrated good activity towards Gram-positive bacteria (e.g. *S. aureus* and methicillin-resistant *S. aureus*, MRSA), the activity towards Gram-negative species (e.g. *E. coli* and *Pseudomonas aeruginosa*) is significantly lower (Poynton et al., 2017; Li et al., 2015). It is worth noting that Gram-negative bacteria with multidrug-resistance are rapidly spreading globally and frequently associated with different surgical and non-surgical infections (Zowawi et al., 2015).

Most ruthenium-based complexes with good therapeutic performance are discrete coordination compounds, in which the antimicrobial activity is associated with their lipophilicity, and charge-related effects, properties that can be modulated by ligands. However, most ligands are toxic, air- and moisture-sensitive and impart poor water solubility (Notaro and Gasser, 2017). Although some studies have been done on the physical encapsulation of Ru complexes in polymeric carriers, or the covalent conjugation of Ru complex to a polymer to promote the release of active Ru drugs (Villemin et al., 2019), up to now, all studies have been focused on carbon-based polymers and Ru metal complexes. Noticeably, the antibacterial properties of Ru based polymeric precursors have not been studied yet. The present investigation offers a facile synthesis of a polymeric Ru-precursor to be used as a photoactivated antimicrobial agent. In particular, the antimicrobial efficiency of the polymeric dicarbonyldichlororuthenium (II) has been proved against *S. aureus* and *E. coli* strains by performing inhibition zone measurements, colony forming capacity, bacterial viability and oxidative stress-induced cell damage. The cytotoxicity profile of this polymeric compound has also been assessed using human dermal fibroblasts, HeLa cells and human red blood cells.

## 2. Materials and Methods

### 2.1. Materials

Formic acid (90% solution, paraformaldehyde) and $RuCl_3 \cdot xH_2O$ were purchased from Sigma Aldrich (now part of Merck Group) and Strem. All the chemicals used for culture media were microbiological grade obtained from Laboratorios Conda (Torrejón de Ardoz, Spain). All other chemical and biochemical reagents were of analytical grade and purchased from Merck (Germany). Ultrapure Millipore water (resistivity >18.2 MΩ) was used as solvent.

### 2.2. Synthesis of dicarbonyldichlororuthenium (II) $[Ru(CO)_2Cl_2]_n$

The polymeric precursor dicarbonyldichlororuthenium (II), $[Ru(CO)_2Cl_2]_n$ was prepared through the $RuCl_3$-assisted decarbonylation of formic acid in the presence of formaldehyde following a method published elsewhere with some modifications (Anderson et al., 1995a; Zeng et al., 2015). Typically, 0.5 g of paraformaldehyde and 1.0 g of $RuCl_3 \cdot xH_2O$ were added to an Ar-purged solution of 90% formic acid (30 mL) and heated under reflux for 6 h. A characteristic colour change was observed from red to green to orange to pale yellow. The reaction vessel was then cooled to room temperature and stored at 4 °C overnight to allow complete conversion. The solution was evaporated to dryness on a steam bath and the residue mixed with hexane, grounded, and vacuum dried. Successive recrystallizations from acetone/diethyl ether yielded the pure polymer. IR ν (CO)$_{st}$ = 2068, 1995 cm$^{-1}$.

### 2.3. Characterization of dicarbonyldichlororuthenium (II), $[Ru(CO)_2Cl_2]_n$

Attenuated Total Reflectance Fourier Transform Infrared spectra (ATR-FTIR) were obtained using a Thermo-Scientific Nicolet iS10 apparatus with 4 cm$^{-1}$ resolution and 1000–4000 cm$^{-1}$ scan range. Thermogravimetric analysis (TGA) of the powder sample was done under argon flow at a heating rate of 10 °C min$^{-1}$ on a simultaneous thermal analyzer (STA 449 F, Netzsch, Germany). Zeta potential of polymeric $[Ru(CO)_2Cl_2]_n$ was determined by electrophoretic light scattering using a Zetasizer Nano-ZS apparatus (Malvern Instruments Ltd. UK). The results are expressed as mean values of three samples. The spectroscopic analysis of polymeric $[Ru(CO)_2Cl_2]_n$ was performed using an ultraviolet-visible spectrophotometer (Cary Varian 50 scan UV–Vis equipped with Cary Win UV software). For XPS study, polymeric $[Ru(CO)_2Cl_2]_n$ was prepared in pellet form and the analysis carried out in a PHI 5000 VersaProbe II spectrophotometer (Physical Electronics Inc., USA) using a monochromatized Al K$_\alpha$ (~1486.6 eV) X-ray beam of size ~ 100 μm. Recorded high resolution C 1s photoelectron spectra were resolved into their respective Gaussian fits after removal of background intensity. Cyclic voltammetry measurements were performed by using a galvanostat-potentiostat (PGSTAT 30, Autolab, Metrohm, Switzerland) electrochemical workstation at a scanning rate of 50 mVs$^{-1}$ with a standard three-electrode electrolytic cell. A Pt foil served as the counter electrode. All potentials were reported against Ag/AgCl reference electrode. A thin film of as synthesized material on glassy carbon electrode was used as working electrode. Linear sweep voltammetry scans have been measured under dark and light illumination from a 250 W xenon arc lamp having output illumination irradiation of 100 mW cm$^{-2}$.

### 2.4. Antimicrobial activity tests

The microorganisms were preserved at −80 °C in glycerol (20% v/v) until use. The microorganisms were reactivated in nutrient broth (NB, 10 g L$^{-1}$ peptone, 5 g L$^{-1}$ sodium chloride, 5 g L$^{-1}$ meat extract and, for solid media, 15 g L$^{-1}$ powder agar, pH 7.0 ± 0.1) by incubation at 37 °C under shaking at 100 rpm. Inoculums were prepared in 1/500 NB with $10^8$ cells mL$^{-1}$ for *E. coli* and $10^{10}$ cells mL$^{-1}$ for *S. aureus* (followed by optical density at 600 nm) ensuring the exponential growth phase of both microorganisms during contact experiments. The antimicrobial and antifouling effects of material were evaluated according to the standardized ISO 22196 test, followed with minor modifications.

The antimicrobial activity of polymeric [Ru(CO)$_2$Cl$_2$]$_n$ was tested against the bacteria *E. coli* (CET 516, coincident with ATCC 8739) and *S. aureus* (CETC 240, coincident with ATCC 6538 P). The minimum inhibitory concentration (MIC) was determined by measuring colony forming units (CFU). Additional details on the manipulation of microorganisms are given in the Supplementary Material. For agar diffusion tests, approx. 5 mg of [Ru(CO)$_2$Cl$_2$]$_n$ was placed on agar plates inoculated with 0.4 mL bacterial suspension with ∼$10^8$ CFU/mL and incubated at 37 °C for 24 h. Inhibition zones were recorded and measured using photographs. Control experiments were run in parallel without the tested material. Liquid incubation tests were carried out for [Ru(CO)$_2$Cl$_2$]$_n$ material both in suspensions and thin films of the compound deposited on glass coverslips. The antibacterial effect of [Ru(CO)$_2$Cl$_2$]$_n$ precursor was assayed in experiments with inoculums of 2.4 mL of $10^8$ cells mL$^{-1}$ using 24-well plates, at different concentrations (0, 3.3, 6.6, 33, 66, 333, 667 ng/mL). Then, aliquots of suspensions were placed in sterile 96 well microliter plates in 10-fold serial dilutions in phosphate buffered saline (PBS). 10 μL spots of each dilution were placed on Petri dishes with NB medium, incubated for 24 h at 37 °C, and CFU counted. At least three replicates from at least two dilutions were used to derive quantitative estimation of CFU inhibition, which were generally expressed as the logarithm of CFU mL$^{-1}$.

Antifouling experiments were performed using films of the polymeric [Ru(CO)$_2$Cl$_2$]$_n$ material coated on glass coverslips. Cover glasses with films of polymeric [Ru(CO)$_2$Cl$_2$]$_n$, and blank coverslips were placed with the functionalized layer facing up into 24-well microplates and put in contact with 2.4 mL of bacterial suspensions as described before. Microplates were incubated at 37 °C for 20 h without stirring. [Ru(CO)$_2$Cl$_2$]$_n$ films were prepared on glass coverslips by dissolving 0, 1, 5 or 10 μg of polymeric [Ru(CO)$_2$Cl$_2$]$_n$ in 1 mL of ethanol dried at room temperature for 12 h.

After incubation with cultures of exponentially growing bacteria, the specimens were irradiated with a 365 nm UV-LED (LED BLS 13000-1, Mightex). LED irradiance was 110.5 mW cm$^{-2}$ and two scenarios were simulated corresponding to Winter-Fall, L(+), and Summer-Spring, L(++), exposure conditions using irradiances of 1.0 kW-h m$^{-2}$ for L(+) and 3.0 kW-h m$^{-2}$ for L(++), which were based on data available from NASA Surface Meteorology and Solar Energy Database. Additional details are included as Supplementary Material. The antibacterial effect of surface deposited polymeric [Ru(CO)$_2$Cl$_2$]$_n$ was quantified by determining CFU in both dark and irradiated experiments as explained before.

### 2.5. Cytotoxicity test

The cytotoxicity of polymeric [Ru(CO)$_2$Cl$_2$]$_n$ was determined by using human dermal fibroblasts (hDF, ATCC) and immortal HeLa cells. Cells were harvested from culture flasks by trypsinization and aliquots of 100 μL seeded in 96-well microplates in densities of $1 \times 10^4$ cells per well and incubated for 24 h at 37 °C in a humidified atmosphere of 5% CO$_2$ in air (approx. 70-80% confluence). Cell toxicity was studied by means of the colorimetric MTT (3-(4,5-dimethyl-2-thiazolyl)-2,5-diphenyl-2H-tetrazolium bromide) assay based on the reduction of tetrazolium salt by the mitochondrial dehydrogenases of viable cells to yield formazan as a coloured insoluble product. The reduction of absorbance can be attributed to a lower number of viable cells or the inhibition of cell proliferation upon exposure to [Ru(CO)$_2$Cl$_2$]$_n$. Stock solutions of [Ru(CO)$_2$Cl$_2$]$_n$ were prepared in PBS, diluted in MEM media and dispensed into wells. After 24 h exposure, $[Ru(CO)_2Cl_2]_n$ was removed, replaced with MEM/MTT mixture and incubated for 4 h. After that, formazan crystals were dissolved in DMSO and absorbance recorded at 570 and 630 nm using a BioTek® Elisa Reader. At least two independent runs with at least three replicas were used for each concentration level. Cell viability was derived from the formazan generated, the amount of which is proportional to the number of metabolically active cells. Untreated cells and media alone were taken as positive and negative controls. The number of surviving cells was expressed as percent viability and calculated as follows:

$$Percent\ viability = 100\ \frac{Sample\ absorbance\ (treated\ cells) - background}{Control\ absorbance\ (untreated\ cells)}$$

The toxicity to human red blood cells (RBCs) was assessed by a haemoglobin release assay as reported elsewhere (Ghosh et al., 2014a). Briefly, the erythrocytes obtained from blood cells were PBS washed (pH 7.2) and resuspended in PBS. Then 100 μL of the erythrocyte solution were incubated with polymeric $[Ru(CO)_2Cl_2]_n$ for 2 h in 96-well plates with PBS. Intact erythrocytes were pelleted by centrifugation at 1000×g for 5 min at 4 °C and the release of haemoglobin was measured using UV–Vis absorbance at 450 nm. The negative and positive controls were PBS and control 1% TritonX-100, respectively. The percent of haemolysis was calculated as follows:

$$Hemolysis\ (\%) = 100\ \frac{Sample\ absorbance - negative\ control}{Positive\ control - negative\ control}$$

### 2.6. Bacterial viability assays

The viability of bacterial cells was assessed using fluorescein diacetate (FDA). FDA is hydrolysed in functional cells by intracellular esterases yielding the fluorescent compound fluorescein. Polymeric $[Ru(CO)_2Cl_2]_n$ was put in contact with 5 μL FDA (0.02 % w/w in dimethylsulfoxide, DMSO) and 195 μL of bacterial suspension in wells of 96-well black microplates, subsequently incubated at 25 °C for 30 min. Fluorescence was then measured every 5 min (Ex. 485 nm; Em. 528 nm) in a Fluoroskan Ascent FL fluorometer (Thermo Scientific). Each sample was measured four times, and viability values were compared with negative controls.

The metabolic activity of bacterial cells in contact with $[Ru(CO)_2Cl_2]_n$ was also assessed using Alamar Blue stain. In brief, samples of $[Ru(CO)_2Cl_2]_n$ suspensions were put in contact with Alamar Blue into 96-well plates and incubated at 37 °C for 1 h. Fluorescence readings were performed every 5 minutes for 60 min (Ex. 530 nm; Em. 590 nm) using the same Fluoroskan Ascent FL fluorometer.

### 2.7. Generation of reactive oxygen species

Intracellular oxidative stress was assessed using 2′,7′-dichlorodihydrofluorescein diacetate ($H_2DCFDA$), which is hydrolysed by intracellular esterases to dichlorofluorescein (DCF). DCF yields a fluorescent compound when oxidized by hydrogen peroxide and other reactive oxygen species (ROS). Briefly, polymeric $[Ru(CO)_2Cl_2]_n$ was put in contact with bacteria cells (150 μL) and incubated with 50 μL of 10 mM 2′,7′-dichlorodihydrofluorescein diacetate ($H_2DCFDA$) in black 96-well plates at 25 °C for 5 min. Fluorescence was recorded every 5 min for 25 min (Ex. 485 nm; Em. 528 nm).

### 2.8. Microscopic studies

The bacteria on functionalized surfaces were visualized using Scanning Electron Microscopy (DSM-950 Zeiss, Oberkochen, Germany). Confocal laser scanning microscopy (CLSM) was performed using a LeicaMicrosys-tems Confocal SP5 Fluorescence microscope (Germany). Cell viability and membrane damage as well as the antifouling capacity were assessed using Live/Dead BacLight Bacterial Viability Kit (Fisher Scientific) as indicated by the producer. This method differently stains viable and non-viable cells using the nuclear acid stains SYTO9 and propidium iodide (PI). SYTO9 marks all cells with green fluorescence, while PI red-marks only membrane-damaged cells.

### 2.9. Statistics

A one-way ANOVA coupled with Tukey's HSD (honestly significant difference) post-hoc test was performed for comparison of means. Statistically significant differences were considered to exist when p-value < 0.05.

### 3. Results and discussion

*3.1. Synthesis and characterization of polymeric $[Ru^{II}(CO)_2Cl_2]_n$*

Polymeric $[Ru^{II}(CO)_2Cl_2]_n$ was prepared through $RuCl_3$-assisted decarbonylation of formic acid in the presence of formaldehyde. The reaction is initiated with the reduction of Ru(III) to Ru (II) promoted by formaldehyde (Eqs. 3-5), it has been proposed that the presence of formaldehyde favours the formation of the polymer (Anderson et al., 1995b; Kuramochi et al., 2015; Cleare and Griffith, 1969). In the reaction a change in colour from orange-brown to pale yellow, is observed. The intermediate green solution may correspond to $\{Ru(CO)(H_2O)Cl_4\}^{2-}$ which might also act as a catalyst (Halpern et al., 1966).

$$RuCl_3 \cdot xH_2O \xrightarrow[\Delta]{HCO_2H, (H_2CO)_n} [Ru(CO)_2Cl_2]_n \quad (3)$$

The possible redox processes involved in the formation of the product are:

Reduction: $RuCl_3 \cdot xH_2O + 2HCOOH + e^- \rightarrow 1/n [Ru(CO)_2Cl_2]_n + 2H_2O + Cl^- + xH_2O$ (4)

Oxidation: $HCOOH + 1/n (H_2CO)_n \rightarrow 2CO + H_2O + 2H^+ + 2e^-$ (5)

Overall: $2 RuCl_3 \cdot xH_2O + 5 HCOOH + 1/n (H_2CO)_n \rightarrow 2n [Ru(CO)_2Cl_2]_n + 5 H_2O + 2 HCl + 2 CO + xH_2O$ (6)

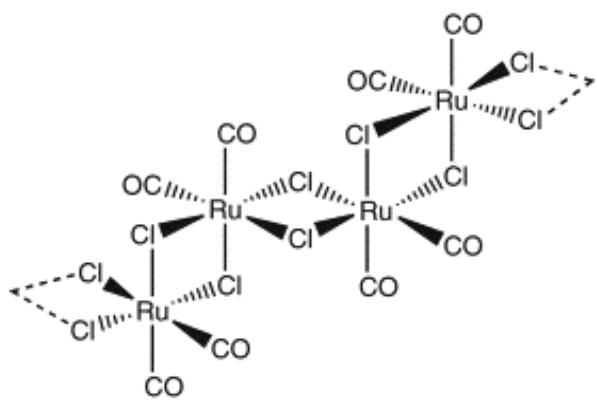

**Fig.1.** Proposed "kinked-chain" structure of $[Ru(CO)_2Cl_2]_n$.

As shown Fig. 1 the proposed structure of the $[Ru(CO)_2Cl_2]_n$ polymer entails a "kinked-chain" with the chloride ligands bridging the metals in a *cis-* and *trans-* disposition, while the carbonyl groups are coordinated as *cis*-terminal ligands, this structure is analogous to the one suggested for $[Mo(NO)_2Cl_2]_n$ (Cotton and Johnson, 1964).

$[Ru(CO)_2Cl_2]_n$ has been characterized using IR spectroscopy since this technique is considered to be the most informative one for metal carbonyls (Nakamoto, 2008; Ziegler et al., 1987). Uncoordinated carbon monoxide displays a band at 2143 cm$^{-1}$ assigned to the C–O bond stretching mode. When coordinated this band shifts to lower values and its position correlates inversely to the π-bonding strength between the metal and carbon. The number and intensity of carbonyl stretching bands largely depend of the local symmetry. The IR spectrum of the polymeric $[Ru^{II}(CO)_2Cl_2]_n$ shown bands at 2068, and 1995 cm$^{-1}$, which can be assigned to the asymmetric and symmetric CO stretching vibrations of the terminal carbonyl groups, which is consistent with reported literature (Fig. 2a) (Anderson et al., 1995b).

The thermal study of the polymeric $[Ru^{II}(CO)_2Cl_2]_n$ was performed with the thermogravimetric technique at heating rate of 10°C min$^{-1}$. Polymeric $[Ru^{II}(CO)_2Cl_2]_n$ decomposes in two steps as shown in Fig. 2b, the first one takes place in the temperature range 70–300 °C with a total mass loss of ~12% which can be attributed to one CO group removal. The further decomposition step (186–620 °C) with a mass loss of ~48% is assigned to the elimination of the other groups leaving RuO as the metallic residue.

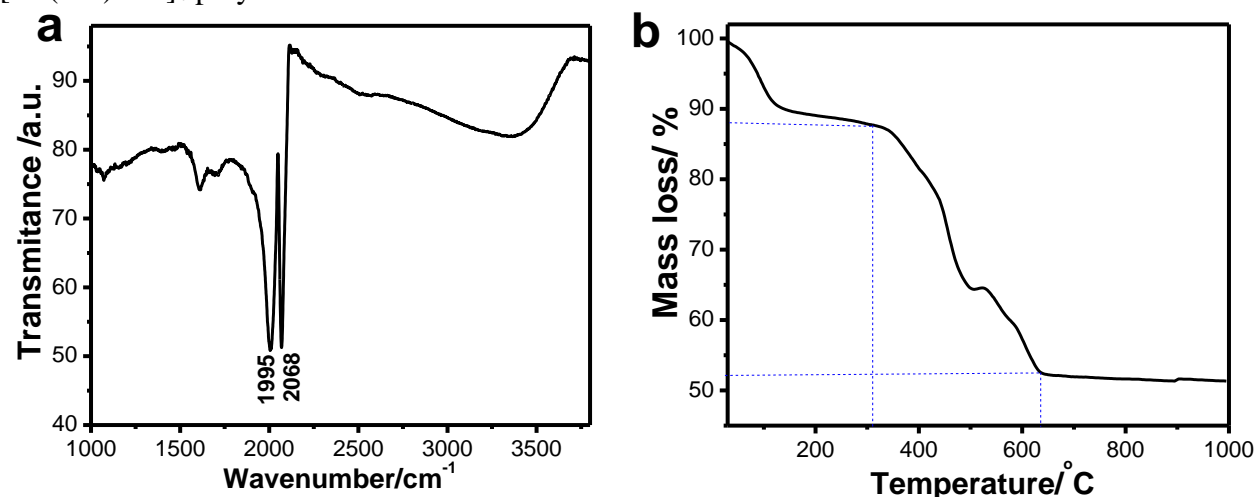

**Fig. 2**. (a) FTIR spectra and (b) TGA profile of Polymeric $[Ru^{II}(CO)_2Cl_2]_n$.

XPS analysis was further performed to study the chemical composition of [Ru(CO)$_2$Cl$_2$]$_n$. Fig. 3a shows the scan survey XPS spectrum that confirms the existence of ruthenium. The Ru 3d XPS signal can be deconvoluted into two distinguishable doublets with different intensities. The doublet with peaks at 281.4 (assigned to Ru 3d$_{5/2}$) and 286.0 eV (assigned to Ru 3d$_{3/2}$) can be indexed to the Ru$^{II}$ state (Zubavichus et al., 2002; Grelaud et al., 2014). The Ru 3d$_{5/2}$ showed two signals at 281.4 and 282.3 eV. The signal at 281.4 eV is characteristic of Ru$^{2+}$ complexes, indicating the formation of Ru(II)-centred polymeric precursors (Fig. 3b). Moreover, the separation distance between the Ru 3d$_{5/2}$ and Ru 3d$_{3/2}$ peaks was 4.0–4.1 eV. Therefore, the Ru 3d$_{3/2}$ line, which could be expected to fall in the range of 284.1–284.2 eV, close to the C 1s line (284.6 eV) may prevent the accurate analysis of the Ru-oxidation state from this binding energy value. In addition, chlorine peaks were observed at 198.7 eV (Fig. 3c), which confirms the presence of Cl in the polymeric [Ru(CO)$_2$Cl$_2$]$_n$ samples (Suñol et al., 2000).

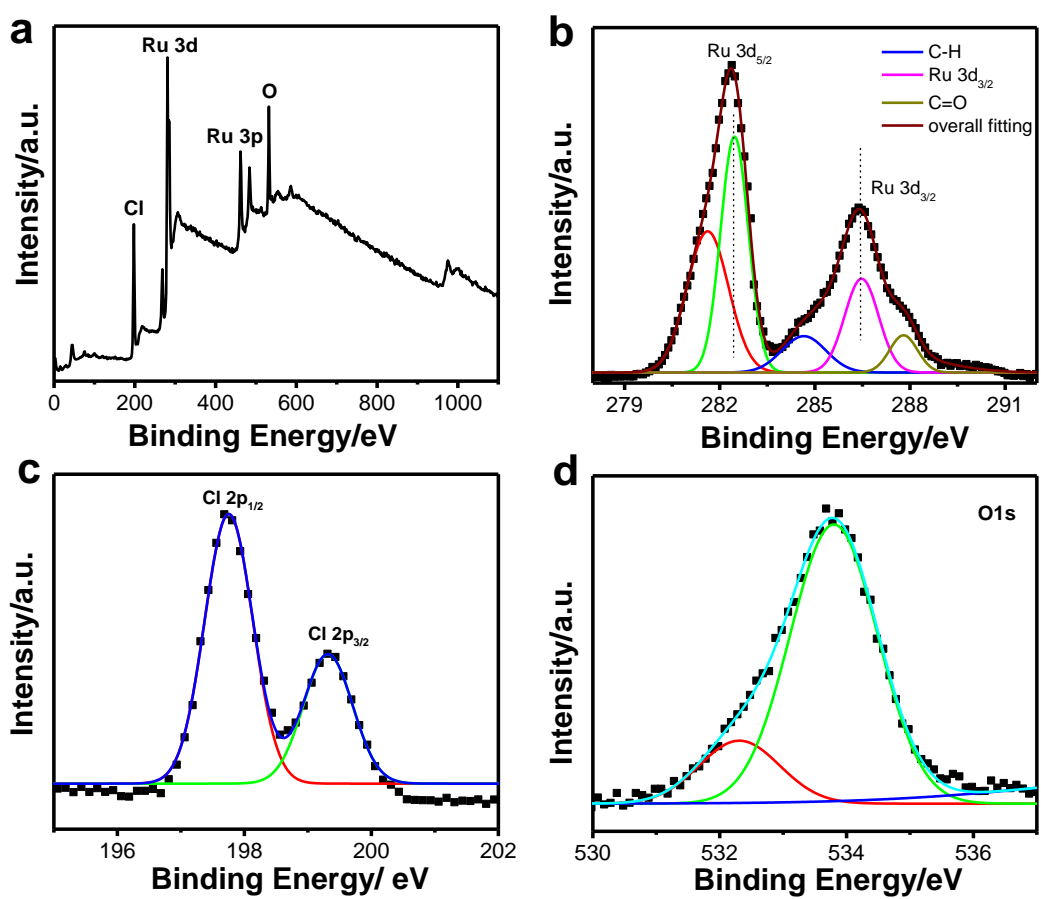

**Fig. 3.** (a) XPS scan survey of [Ru(CO)$_2$Cl$_2$]$_n$ and enlarged spectra of (b) Ru 3d, (c) Cl 2p, (d) O 1s.

The photophysical properties of polymeric [Ru(CO)$_2$Cl$_2$]$_n$ were investigated with spectroscopic measurements. The Fig. 4a shows the UV-Vis absorption spectra of polymeric [Ru(CO)$_2$Cl$_2$]$_n$ and displays two absorption bands at 370 and 245 nm. The absorptions in the region *ca.* 320–400 nm is due to the metal ligand charge transfer (MLCT) transitions primarily, while the more intense bands to higher energies in the UV region are attributable to $\pi \rightarrow \pi^*$ transitions (Roy et al., 2017). Hence, polymeric [Ru(CO)$_2$Cl$_2$]$_n$ showed strong absorption intensity in the visible range region. Upon excitation at 320 nm, shows a strong emission with a maximum around 445 nm as shown in Fig. 4b. Further, the stability of polymeric [Ru$^{II}$(CO)$_2$Cl$_2$]$_n$ was tested under UV irradiation up to 240 mins and compared the absorption spectra before and after the treatment. There was no significant change observed in the absorption profile and the absorbance value monitored at $\lambda_{370}$ nm, which showed 2% lowering in the absorbance value after 240 mins irradiation with UV light. This suggests high stability of polymeric [Ru$^{II}$(CO)$_2$Cl$_2$]$_n$ under UV irradiation.

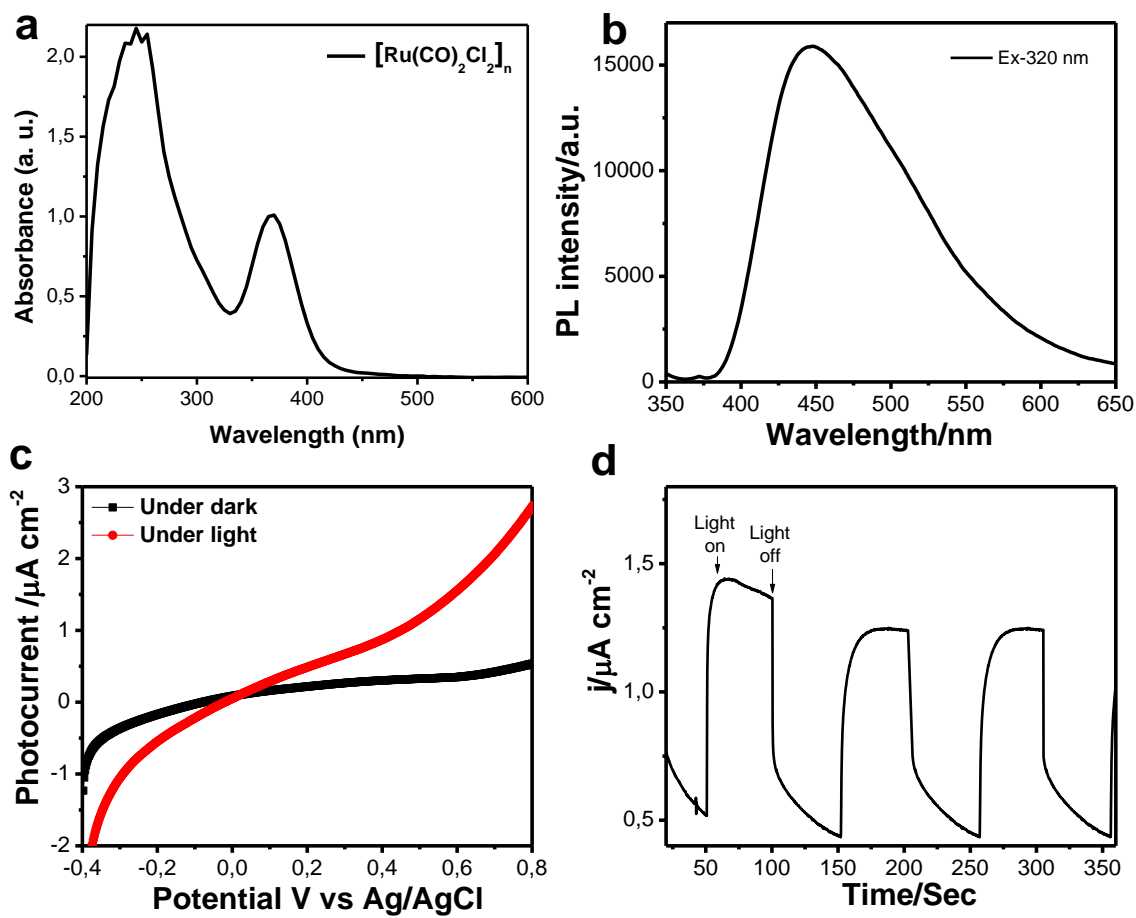

**Fig. 4.** (a) UV−vis spectrum and (b) photoluminescence spectra of polymeric [Ru(CO)$_2$Cl$_2$]n. (c) LSV scans for polymeric [Ru(CO)$_2$Cl$_2$]n in 0.1 M Na$_2$SO$_4$ solution both under dark and under light illumination with output intensity 100 mW cm$^{-2}$. The scan rate was 20 mVs$^{-1}$. (d) Photocurrent response of polymeric [Ru(CO)$_2$Cl$_2$]$_n$ at zero bias voltage under UV light irradiation.

Furthermore, the photoelectrochemical properties of polymeric [Ru(CO)$_2$Cl$_2$]$_n$ have been investigated by linear-sweep-voltammetry (LSV) measurements using a three electrode photoelectrochemical (PEC) cell in 0.1 M Na$_2$SO$_4$ solution both in dark and under irradiation (Fig. 4c). The dark current for polymeric [Ru(CO)$_2$Cl$_2$]$_n$ is nearly zero from 0 V to 0.8 V (vs. Ag/AgCl) whereas the photocurrent density reached a value of 2.8 μA cm$^{-2}$ at a potential of 0.8 V vs. Ag/AgCl reference electrode under light irradiation. Fig. 4d shows the photocurrent response of polymeric [Ru(CO)$_2$Cl$_2$]$_n$ during repetitive on/off cycles of light illumination. The repetitive on−off response to ultraviolet light (100 Wcm$^{-2}$ at 254 nm) was found to be stable and reversible. This suggested an active photo response of [Ru(CO)$_2$Cl$_2$]$_n$ suitable for photocatalytic applications.

### 3.2. Antibacterial effect of [Ru(CO)$_2$Cl$_2$]$_n$

The antimicrobial effect of the polymeric [Ru(CO)$_2$Cl$_2$]$_n$ was investigated against *E. coli* and *S. aureus*, the two most frequently encountered pathogens. Agar diffusion tests showed clear inhibition of bacterial growth, as shown in Fig. 5, indicating that polymeric [Ru(CO)$_2$Cl$_2$]$_n$ is effective against both bacterial strains, thereby demonstrating a broad spectrum of antimicrobial activity.

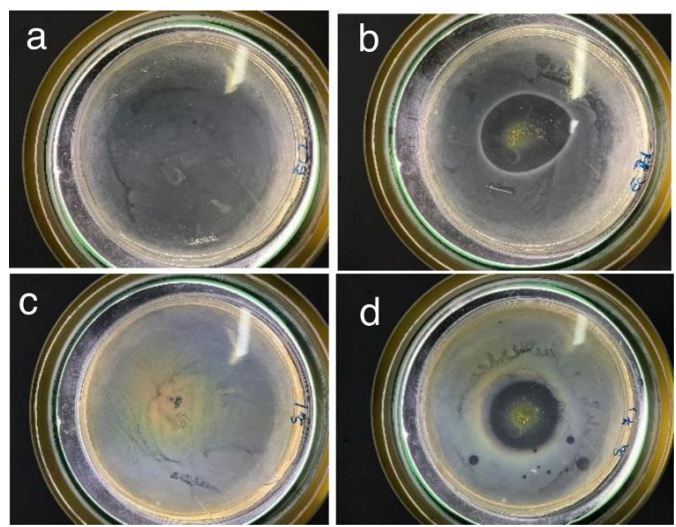

**Fig. 5.** Antimicrobial properties of the polymeric [Ru(CO)$_2$Cl$_2$]$_n$. Photographs of (a) control *E. coli,* (b) *E. coli* treated with [Ru(CO)$_2$Cl$_2$]$_n$ and (c) control *S. aureus* and (d) *S. aureus* treated with polymeric [Ru(CO)$_2$Cl$_2$]$_n$ after 24 h.

To quantify the antibacterial effect of polymeric [Ru(CO)$_2$Cl$_2$]$_n$, liquid incubation tests with quantitative CFU measurements were performed as indicated before. Fig. 6 displays the inactivation of *E. coli* and *S. aureus* upon exposure to [Ru(CO)$_2$Cl$_2$]$_n$ under different light conditions: dark, L(-), Winter-Fall, L(+), and Summer-Spring, L(++), based on 365 nm exposure to UV-A irradiation. In the absence of irradiation, [Ru(CO)$_2$Cl$_2$]$_n$ did not inhibit bacterial growth below 33 ng/mL. Upon irradiation, *E. coli* growth was inhibited (> 5-log or > 99.999 % inhibition) for [Ru(CO)$_2$Cl$_2$]$_n$ concentration as low 6.6 ng/mL with complete inhibition (no CFU detected) at 33 ng/mL. A similar trend was observed for *S. aureus,* with complete inhibition at 166 ng/mL under L(++) irradiation and > 5-log decrease for [Ru(CO)$_2$Cl$_2$]$_n$ concentration as low 6.6 ng/mL both for L(+) and L(++) assays. Such antibacterial efficiency is not parallel at other ruthenium complexes where antibacterial activity has been reported (Li et al., 2015; Southam et al., 2017).

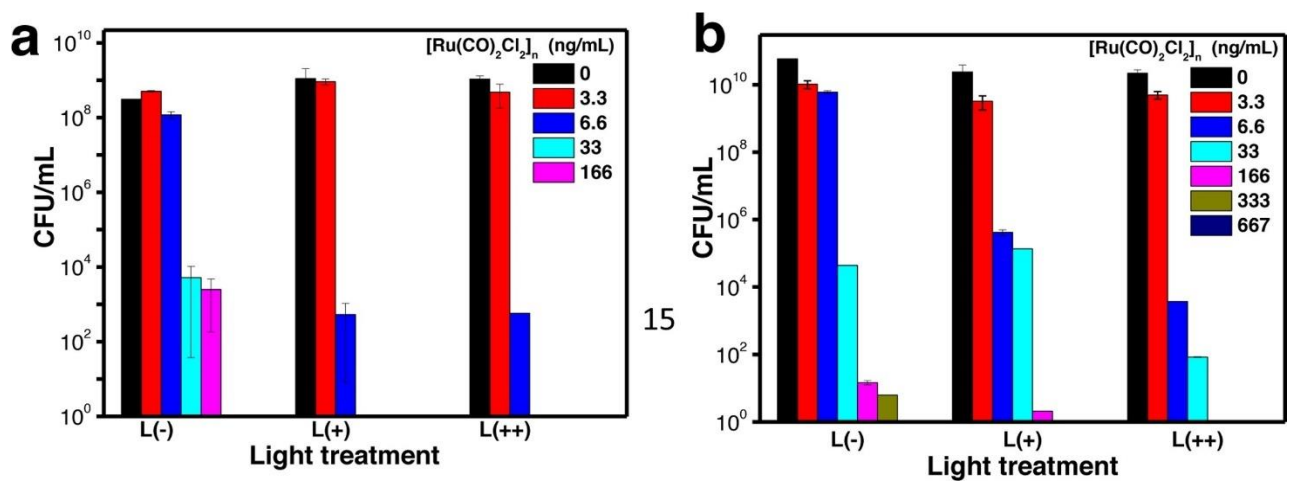

**Fig. 6**. Colony forming units (CFU/mL$^{-1}$) for cultures of (a) *E. coli* and (b) *S. aureus* (represented as the log$_{10}$(CFU mL$^{-1}$)) in contact with suspensions of [Ru(CO)$_2$Cl$_2$]$_n$. The samples were irradiated using a UV lamp, emitting at 365 nm: Dark, L(—), Winter-Fall, L(+), and Summer-Spring, L(++).

Under radiation [Ru(CO)$_2$Cl$_2$]$_n$ showed strong bactericidal activity with minimum inhibitory concentration (MIC) ∼33 ng/mL against Gram-negative bacteria (*E. coli*) and with MIC ∼166 ng/mL, against Gram-positive bacteria (*S. aureus*). This is a remarkable result since Gram-negative bacteria are particularly resistant to stressors due to their additional outer membrane (Gill et al., 2009). The antimicrobial efficiency of [Ru(CO)$_2$Cl$_2$]$_n$ was not the consequence of electrostatic interaction as both polymeric ruthenium complex and pouter bacterial envelopes were negatively charged. The most probable explanation is that the antimicrobial effect is due to the role played by the central Ru atoms (Li et al., 2015).

The biocompatibility and specificity of polymeric [Ru(CO)$_2$Cl$_2$]$_n$ were studied by evaluating its cytotoxicity towards human cell lines, and its haemolytic activity. After treatment with [Ru(CO)$_2$Cl$_2$]$_n$, the viability of both the human dermal fibroblasts and cervical cancer cells (HeLa) was assessed via the MTT cell viability assay. Polymeric [Ru(CO)$_2$Cl$_2$]$_n$ exhibited remarkable biocompatibility, with viability > 95% for human dermal fibroblasts cells at a concentration of 3330 ng/mL, ten times higher than the minimum inhibitory concentration against *E. coli* and *S. aureus* (Fig. 7a).

This suggests that [Ru(CO)$_2$Cl$_2$]$_n$ displays antibacterial activity without impairing human fibroblasts. In contrast, cellular viability of HeLa cells was reduced by ∼50% when treated with 3330 ng/mL of [Ru(CO)$_2$Cl$_2$]$_n$ (Fig.7b). Haemolytic activity was evaluated by quantifying the haemoglobin released from red blood cells upon exposure to [Ru(CO)$_2$Cl$_2$]$_n$. It has been reported that *in vitro* haemolysis from 5 to 25% can be considered as non-toxic level (Dobrovolskaia et al., 2009). Our data showed that [Ru(CO)$_2$Cl$_2$]$_n$ exhibited biocompatibility over a wide concentration range and > 96% of red blood cells remained intact and undisrupted at 3330 ng/mL, a concentration much higher than MIC for bacteria (Fig. 7c). These results indicated that polymeric [Ru(CO)$_2$Cl$_2$]$_n$ was not harmful to non-tumoral cells at bactericidal concentration.

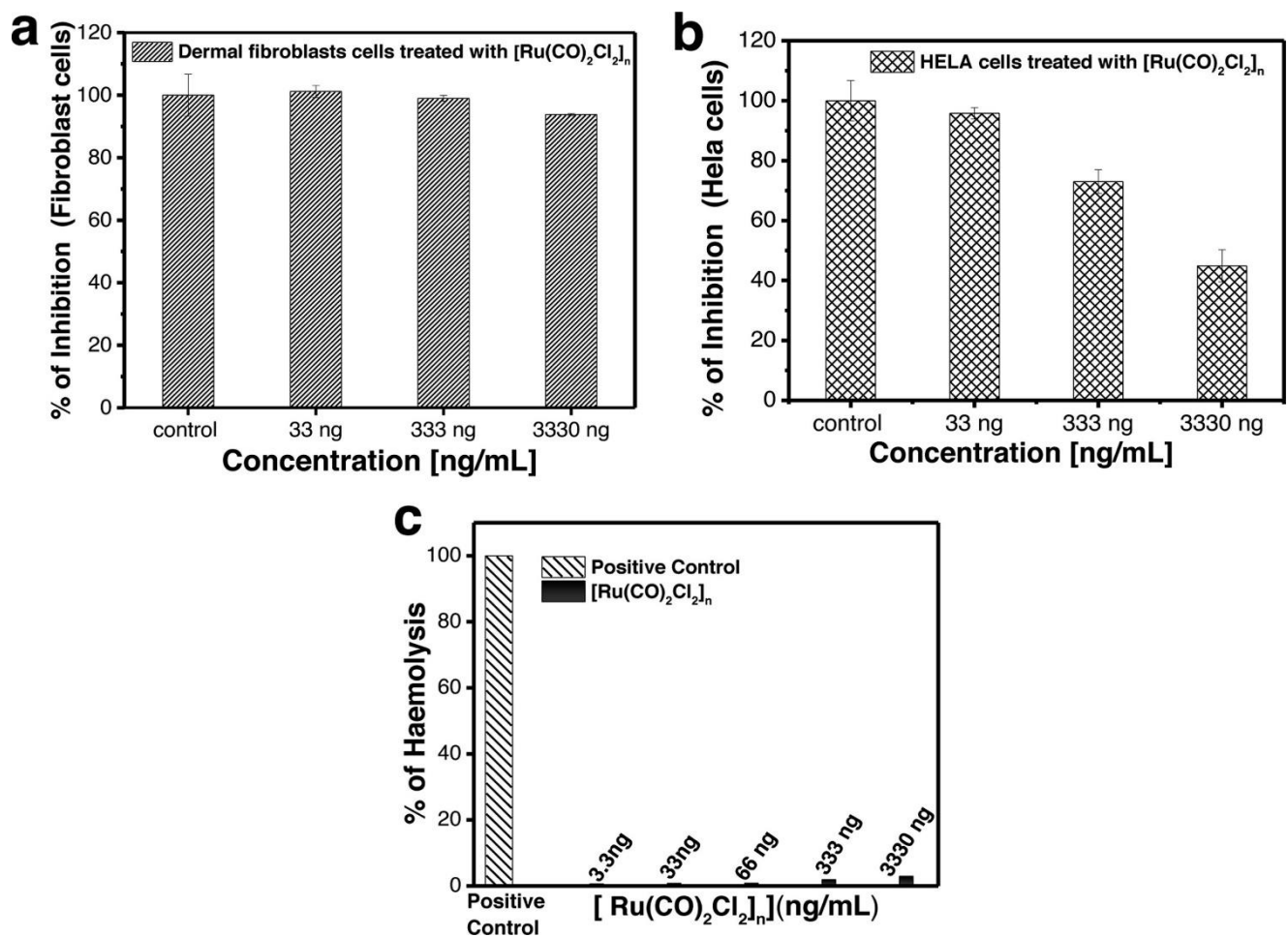

**Fig. 7**. (a) Cell viability of human dermal fibroblasts and (b) HeLa cells after 24 -hs treatment with different concentrations of polymeric $[Ru(CO)_2Cl_2]_n$ as calculated from the MTT assay. (c) Haemolysis assay for polymeric $[Ru(CO)_2Cl_2]_n$ using Triton-X as a positive control. Data are shown as mean ± SD for three independent experiments.

The mechanism effect behind the antibacterial activity of $[Ru(CO)_2Cl_2]_n$ was followed by tracking the metabolic activity of exposed cells and the production of ROS as shown in Fig. 8. The use of fluorogenic substrates provides two indicators of bacterial viability via the detection of cell enzymatic activity. An FDA probe was used as a cell viability indicator through the accumulation of the polar fluorescent compound fluorescein due to the activity of intracellular esterases (Clarke et al., 2001). Polymeric $[Ru(CO)_2Cl_2]_n$ generated significant dose dependent toxicity towards *E. coli*. With increasing concentration of $[Ru(CO)_2Cl_2]_n$, the toxicity effect is more prominent for bacteria. At bactericidal concentration (33 ng/mL), 80% cells were still viable under dark conditions. In contrast, only ∼30% cells were found viable under irradiation, a result coincident with CFU measurements. The effect of increasing $[Ru(CO)_2Cl_2]_n$ concentration was lower for viability than for CFU counts (Fig. 6), which can be attributed to viable but non-culturable bacteria with cells with metabolic activity but unable to replicate. This result was consistent with previously reported literature (Santiago-Morales et al., 2016). Notably, a further increase in $[Ru(CO)_2Cl_2]_n$ concentration, up to 166 ng/mL, resulted in no viable cells under irradiation as shown in Fig. 8a. A similar result was obtained for *S. aureus* (Fig. 8d). Fig. 8b shows the generation of intracellular ROS in *E. coli* under exposure as measured by DCF probe.

Alamar Blue assays showed that up to 6.6 ng/mL, $[Ru(CO)_2Cl_2]_n$ did not influence cellular metabolic activity in the absence of irradiation, with > 90% healthy *E. coli* cells (Rampersad, 2012). However, 50% reduction in metabolic activity was observed under light exposure for the same concentration (Fig. 8c). Besides, significant decrease in metabolic activity was detected following exposure of the cells to 667 ng/mL of $[Ru(CO)_2Cl_2]_n$. A similar trend was detected for ROS generation and Alamar Blue assays, with dose-dependent responses upon $[Ru(CO)_2Cl_2]_n$ exposure, which is particularly evident for *S. aureus* (Fig. 8e-f).

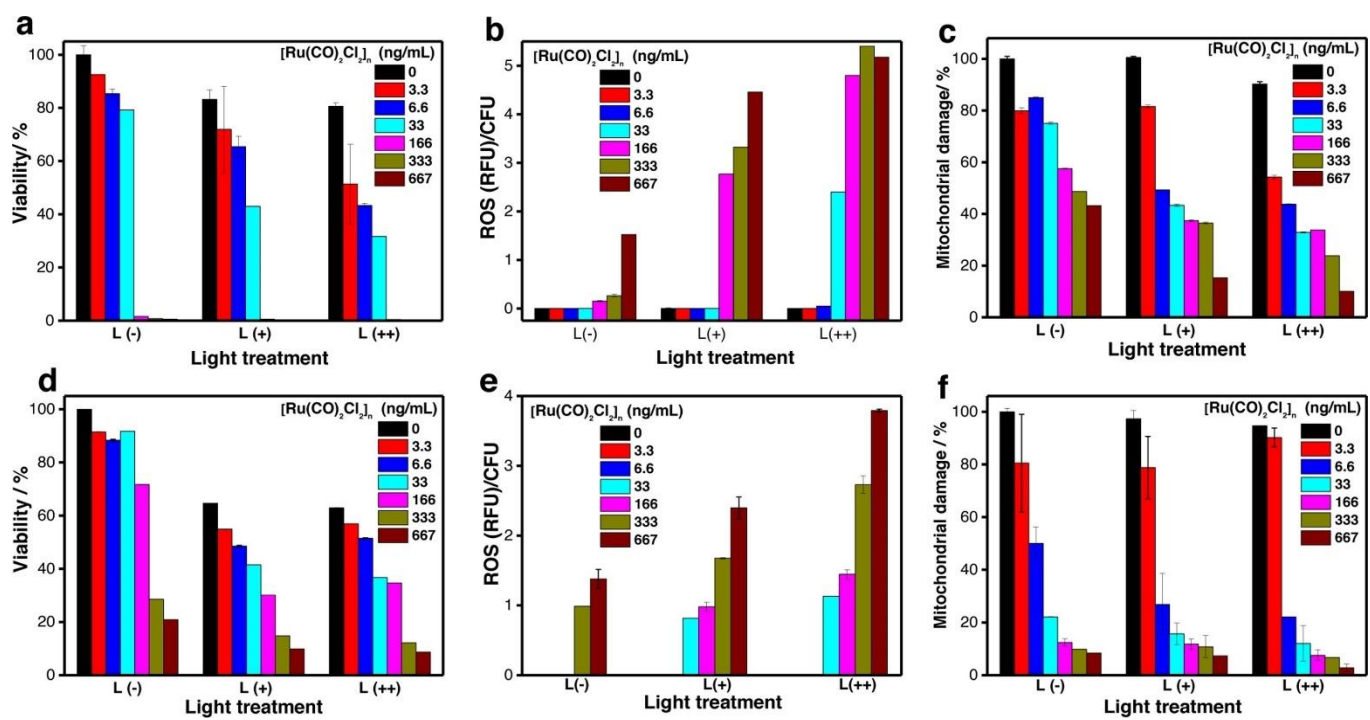

**Fig. 8.** Viability (FDA), ROS (DCF), and metabolic activity (Alamar Blue) assys for *E.coli* (a-c) and *S. aureus* (d-f) kept 20 h in contact with suspensions of polymeric [Ru(CO)$_2$Cl$_2$]$_n$ irradiated using 365 nm. L(-): dark conditions; L(+): 1.0 kW-h m$^{-2}$ (Winter-Fall); L(++): 2.0 kW-h m$^{-2}$ (Summer-Spring).

The experimental results suggest a possible mechanism of the bactericidal activity of polymeric [Ru(CO)$_2$Cl$_2$]$_n$ may involve a ROS-dependent pathway. It is well-known that ROS such as superoxide anions (O$_2^{\bullet-}$), hydrogen peroxide (H$_2$O$_2$) and hydroxyl radicals (OH$^\bullet$) damage lipids, proteins and nucleic acids in cells, in a process that may lead to cell death (Circu and Aw, 2010; Zhao and Drlica, 2014; Hong et al., 2019). In general, metabolic activity involves ROS formation, but healthy cells can efficiently neutralize them. However, when the ROS concentration inside the cell becomes excessive, it can impair the structure and activity of proteins, eventually leading to cell death. Interestingly, recent research indicated that once a ROS threshold is exceeded, ROS accumulation and cell death take place even if the stressor is removed (Ezraty et al., 2017).

### 3.3. Antifouling photo-bactericidal activity of polymeric [Ru(CO)$_2$Cl$_2$]$_n$

The confocal micrographs shown in Fig. 9, Fig. 10 correspond to glass surfaces with and without polymeric coating and three different [Ru(CO)$_2$Cl$_2$]$_n$ concentrations (c1, c2 and c3) in non-irradiated, L(−) and irradiated L(++) samples. *E. coli* and *S. aureus* grow normally in controls, with many green-labelled cells. Live/Dead biofilm staining showed control biofilms compact and with no red-marked cells. Besides, UV irradiation did not induce any bacterial damage in the absence of polymeric [Ru(CO)$_2$Cl$_2$]$_n$. After being treated with [Ru(CO)$_2$Cl$_2$]$_n$ both under dark and UV irradiation, *E. coli* showed cell photo-impairment, with red-marked as well as yellowish or orange-marked cells (Fig. 9).

It has been reported that yellow cells are generally viable cells, while orange cells can be considered damaged (Jalvo et al., 2017). However, under light irradiation, practically all cells were red marked through internalizing PI, which indicates cell membrane damage. Photochemically produced ROS in [Ru(CO)$_2$Cl$_2$]$_n$ functionalized surfaces may result in damage to cell envelopes, eventually leading to cell disorganization and death (Matsumura et al., 2003; Ghosh et al., 2014b).

*S. aureus* biofilms grown on surfaces with [Ru(CO)$_2$Cl$_2$]$_n$ showed viable green-labelled cells with a certain number of cells with compromised membranes (yellow stained) even in dark conditions. Essentially no viable cells were observed in irradiated functionalized surfaces (Fig. 10). This indicates that [Ru(CO)$_2$Cl$_2$]$_n$ treatment damages *S. aureus* in biofilms, which is consistent with quantitative ROS generation data (Fig. 8).

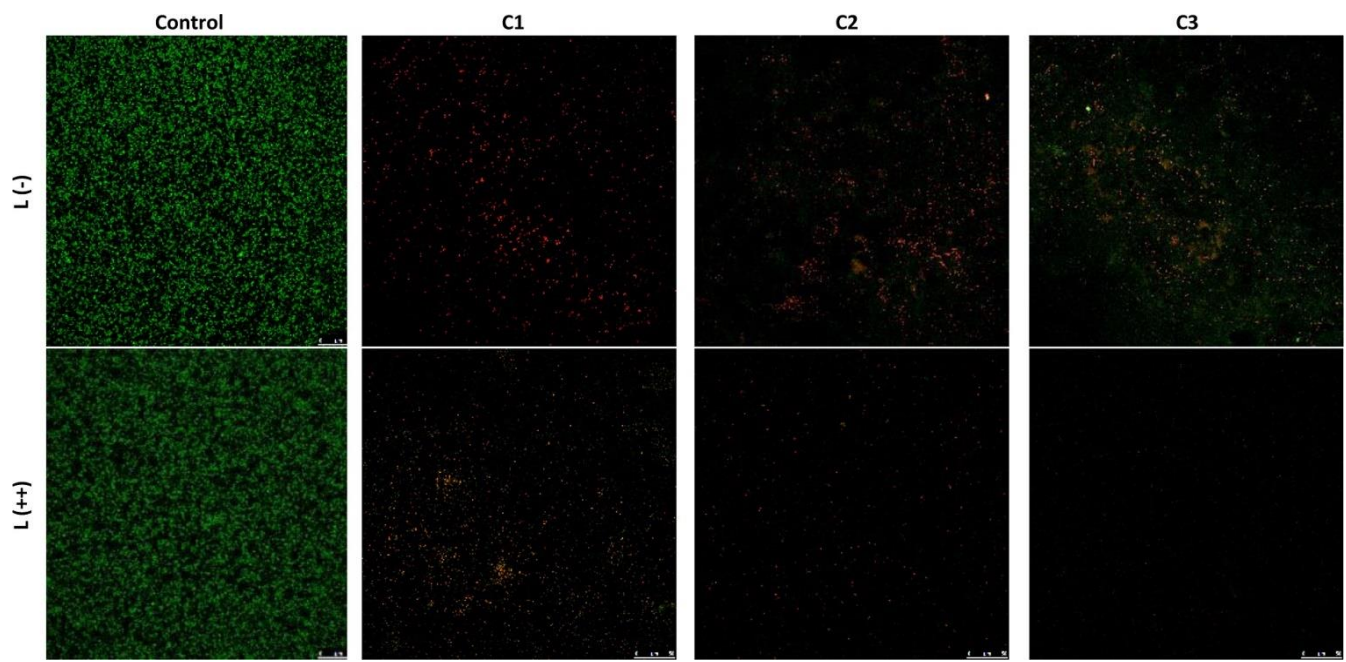

**Fig. 9.** Live/Dead confocal micrographs of *E. coli* biofilm on control glass surface and glass functionalized with polymeric $[Ru(CO)_2Cl_2]_n$ after biofilm growth for 20 h in dark, L(-), and after 20 h in dark plus 1 h UV irradiation, L(++) c1, c2, c3 with $[Ru(CO)_2Cl_2]_n$ concentrations of 0.59, 2.9, 5.9 μg/cm$^2$ respectively. (Scale bar represents 50 μm).

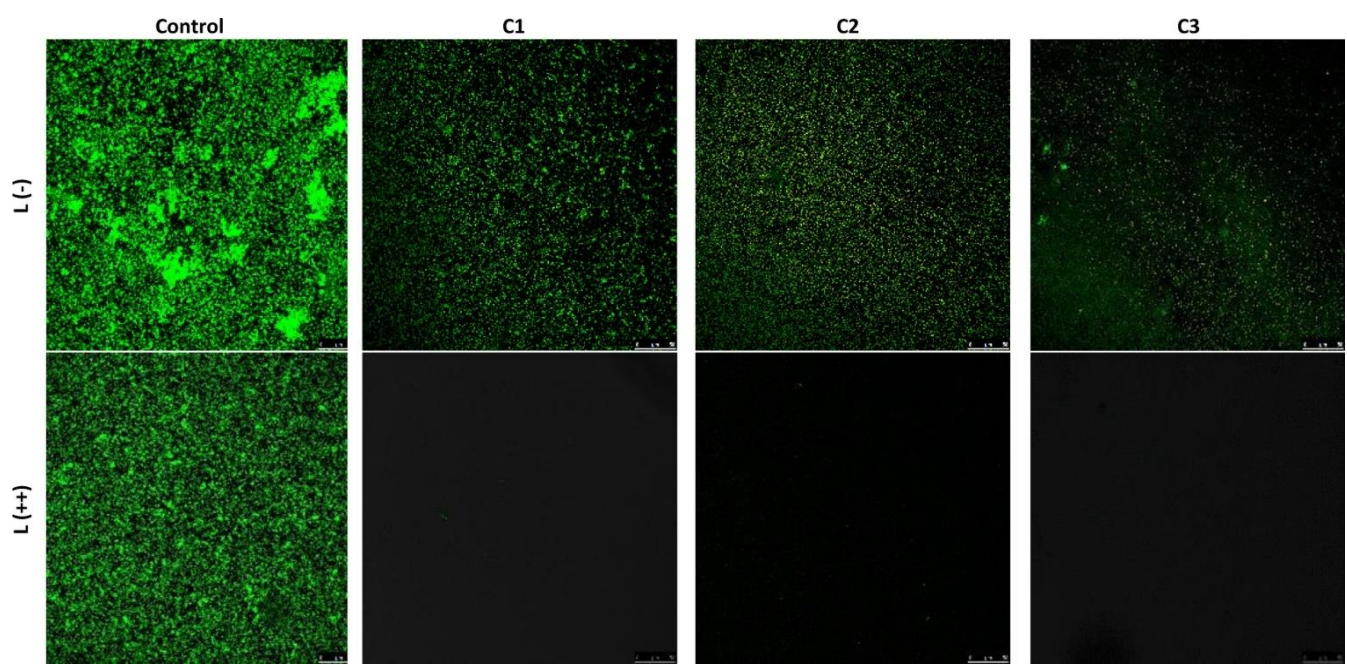

**Fig. 10**. Live/Dead confocal micrographs of *S. aureus* biofilm on control glass surface and glass functionalized with polymeric $[Ru(CO)_2Cl_2]_n$ after biofilm growth for 20 h in dark, L(-), and after 20 h in dark plus 1 h UV irradiation, L(++). Control and c1, c2, c3 with polymeric $[Ru(CO)_2Cl_2]_n$ concentrations of 0.59, 2.9, 5.9 μg/cm$^2$ respectively. (scale bar represents 50 μm).

The changes that occur in biofilm structures upon exposure to $[Ru(CO)_2Cl_2]_n$ were evaluated using SEM. SEM images of *E. coli* and *S. aureus* after exposure are shown in Fig. 11, Fig. 12, respectively. It is worth noting that bacteria in the biofilms are protected by a self-produced matrix consisting of extracellular polymeric substances (EPS) that provide resistance against stress factors, including antibiotics (Donlan and Costerton, 2002; Flemming and Wingender, 2010). SEM showed mature biofilms with their characteristic EPS matrices and extensive bacterial colonization in control samples. SEM imaging was consistent with a dose-dependent reduction in *E. coli* bacterial density and biofilm growth in the presence of $[Ru(CO)_2Cl_2]_n$ (Fig. 11). Bacterial adhesion structures were clearly observed for the lower concentration of $[Ru(CO)_2Cl_2]_n$. However, with further increased concentration of $[Ru(CO)_2Cl_2]_n$, the surfaces were almost free from bacteria under irradiation.

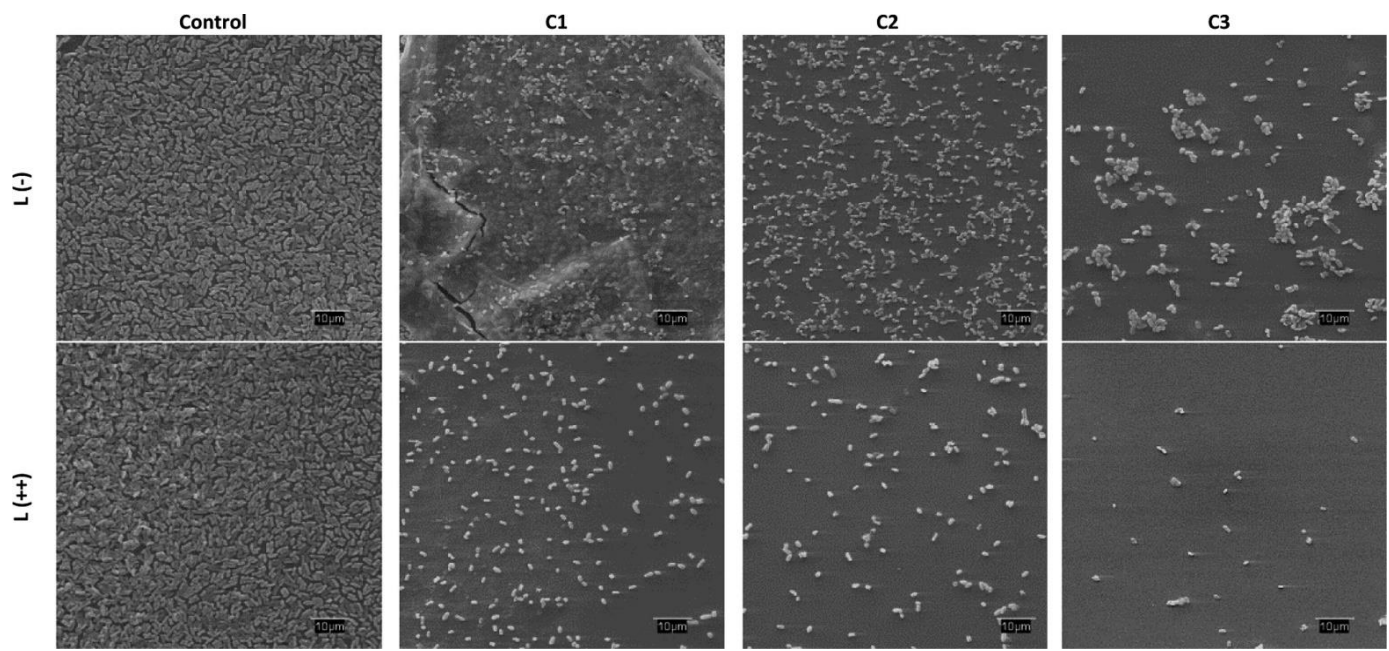

**Fig. 11.** SEM micrographs of *E. coli* biofilm on control glass surface and glass functionalized with polymeric [Ru(CO)$_2$Cl$_2$]$_n$ after biofilm growth for 20 h in dark L(-), and 20 h in dark plus 1 h UV irradiation, L(++). Control and c1, c2, c3 with polymeric [Ru(CO)$_2$Cl$_2$]$_n$ concentrations of 0.59, 2.9, 5.9 μg/cm$^2$, respectively. (scale bar represents 10 μm).

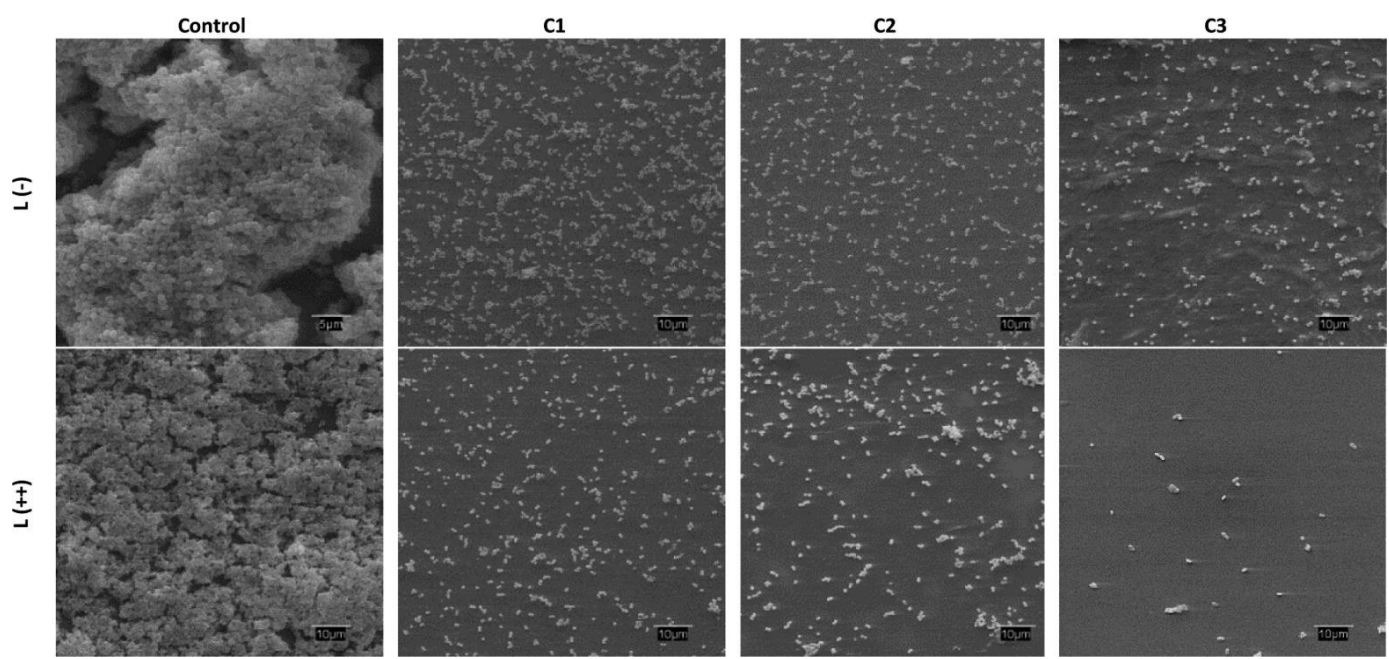

**Fig. 12.** SEM micrographs of *S. aureus* biofilm on control glass surface and glass functionalized with polymeric [Ru(CO)$_2$Cl$_2$]$_n$ after biofilm growth for 20 h in dark L(-), and 20 h in dark plus 1 h UV irradiation, L(++). Control and c1, c2, c3 with polymeric [Ru (CO)$_2$Cl$_2$]$_n$ concentrations of 0.59, 2.9, 5.9 μg/cm$^2$, respectively. (scale bar represents 10 μm).

Similar trend was observed for *S. aureus*. Increasing [Ru(CO)$_2$Cl$_2$]$_n$ concentration resulted in significant removal of bacteria under irradiation. Besides, the biofilms treated with [Ru(CO)$_2$Cl$_2$]$_n$ were thin, and the cells appeared distorted. Hence polymeric [Ru(CO)$_2$Cl$_2$]$_n$ also impaired bacteria within their biofilm and was able to disassemble the biofilm matrix.

Altogether, Live/Dead biofilm staining, and SEM analysis revealed that [Ru(CO)$_2$Cl$_2$]$_n$ altered cell permeability leading to cell lysis. This suggests that bacterial membranes were the target of [Ru(CO)$_2$Cl$_2$]$_n$ activity and due to destabilization of the bacterial wall or the impairment of the lipopolysaccharide interlocking structure, the bacteria become impaired (Ghosh et al., 2016; Arenas-Vivo et al., 2019; Baral et al., 2016). The present study indicates that the antimicrobial polymeric [Ru(CO)$_2$Cl$_2$]$_n$ possess a very high activity with 99.9% removal of bacteria with concentration under light irradiation at nanogram/mL.

**Table 1**. Comparative antibacterial activity (minimum inhibitory concentrations, MIC and minimum bactericidal concentrations, MBC at µg/mL) for the selected ruthenium complexes and polymeric $[Ru(CO)_2Cl_2]_n$ against Gram-positive and Gram-negative bacterial strains.

| Antimicrobial Agent | Bacterial stain | MIC (µg/mL) | MBC (µg/mL) | Reference |
|---|---|---|---|---|
| $[Ru(bb_7)(dppz)]^{2+}$ dppz = dipyrido[3,2-a:2′,3′-c]phenazine; $bb_7$ = bis[4(4'-methyl-2,2'-bipyridyl)]-1,7-alkane | Methicillin-sensitive strain *S. aureus* | 2 | 4 | 57 |
| | *E. coli MG1655* | 8 | 8 | |
| $[Ru(phen)_2(dppz)]^{2+}$ | *S. aureus* | 4 | 8-16 | 57 |
| | *E. coli MG1655* | 64 | 64 | |
| $[Ru(bb_7)(Me_2phen)]^{2+}$ phen = 1,10-phenanthroline and its 5-nitro-, 4,7-dimethyl- and 3,4,7,8-tetramethyl- derivatives | *S. aureus* | 8 | 64 | 58 |
| | *E. coli MG1655* | 32 | ≥128 | |
| $[\{Ru(bpy)_2\}_2(tpphz)]^{4+}$ bpy = 2, 2'-bipyridine | *S. aureus* | 40 µM | 80 µM | 50 |
| $[Ru(2,9-Me_2phen)_2(dppz)]^{2+}$ 2,9-$Me_2$phen = 2,9-dimethyl-1,10-phenanthroline | *S. aureus MSSA160* | 8 | 32 | 60 |
| | *E. coli* | ND | ND | |
| $[Ru(phen)_2dpq]^{2+}$ | *S. aureus MSSA160* | 64 | 128 | 60 |
| $[Ru(bpy)_2dpqC]^{2+}$ | *S. aureus MSSA160* | 32 | 128 | 60 |
| Ru(II) polypyridine with a di-fluorinated dppz under photoactivated | *S. aureus* | 8 µM | 16 µM | 61 |
| | *E. coli* | 16 µM | 32 µM | |
| $[Ru(phen)_2(tip)](ClO_4)_2$ Phen = 1,10-phenanthroline | *S. aureus* | >30 µM | ND | 62 |
| | *E. coli* | >30 µM | ND | |
| $Ru(L)_2bdppz]^{2+}$ L = 2, 20-bipyridine or 1,10-phenanthro-line, bdppz= 9a,13a-dihydro-4,5,9,14-tetraaza-benzotri-phenylene-11-yl)-phenyl-methanone | *S. aureus* | 1500 | | 63 |
| | *E. coli* | 1500 | | |
| $[Ru(phen)_2(G)Cl]_2Cl \cdot H_2O$ 1,10-phenanthroline, guanide | *S. aureus* | 400 | ND | 64 |
| | *E. coli* | 300 | ND | |
| $[Ru(dmob)_3]Cl_2$(dmob = 4,40-dimethoxy-2,20-bipyridine | *S. aureus* | 12.5 | ND | 65 |
| Ruthenium(II) complexes with N-phenyl-substituted 4,5-diazafluorenes, Ru-C7 complex | Methicillin-resistant *S. aureus* (MRSA) | 6.25 | 25 | 66 |
| Ru-C6 complex | MRSA | 25 | >100 | 66 |
| $[Ru(R-pytri)_3]^{2+}(X^-)_2$ complexes, $X^–$ = $PF_6^–$ or $Cl^–$ R-pytri = 2-(1-R-1H-1,2,3-triazol-4-yl)pyridine | *S. aureus* | 1–8 | ND | 67 |
| | *E. coli* | 16-128 | ND | |
| $[\{Ru(tpy)Cl\}_2\{\mu\text{-}bbn\}]^{2/4+}$ Cl-Mbbn; $Rubb_{16}$ tpy = 2,2′:6′,2″-terpyridine; and bbn = bis[4(4′-methyl-2,2′-bipyridyl)]-1, n-alkane (n = 7, 12 or 16) | *S. aureus* | 1 | 1 | 68 |
| | *E. coli* | 2 | 2 | |
| Polymeric $[Ru(CO)_2Cl_2]_n$ | *S. aureus* | 33 ng/mL | 166 ng/mL | This work |
| | *E. coli* | 6.6 ng/mL | 33 ng/mL | |

Note: Data are collected as minimal inhibitory concentrations (MICs) and minimum bactericidal concentrations (MBC) according to the Clinical and Laboratory Standards Institute (CLSI). ND = not determined.

We have compared the antimicrobial activity of the present material with Ru-based antimicrobial from the literature as shown in Table 1. It can be clearly observed from Table 1, that the polymeric [Ru(CO)$_2$Cl$_2$]$_n$ exhibits relatively low minimal inhibitory concentrations (MICs) and minimum bactericidal concentrations (MBC) values compared to traditional Ru-complexes (Liu et al., 2018; Sun et al., 2018; Smitten et al., 2020; Bolhuis et al., 2011; Sun et al., 2020; Sun et al., 2015; Kumar et al., 2009; Abebe and Hailemariam, 2016; Donnelly et al., 2007; Lam et al., 2014; Kumar et al., 2016; Pandrala et al., 2013). We have shown a unique photo-induced bacteria inactivation by using [Ru(CO)$_2$Cl$_2$]$_n$ and a plausible mechanism for the formation of oxidative species for light activated therapy. Taken together, the results presented in this study endorse the significance of [Ru(CO)$_2$Cl$_2$]$_n$ as an antimicrobial agent, and useful for the development of attractive antimicrobial drugs and biomedical materials.

## 4. Conclusions

In summary, we have shown that a simple polymeric ruthenium precursor [Ru(CO)$_2$Cl$_2$]$_n$ is an effective antimicrobial agent for the selective inactivation of pathogenic bacteria under UV-A irradiation. The low MIC values (in the nanogram per millilitre range) obtained in this work suggests that this material has a promising applicability against different types of bacteria. We have also shown that the antimicrobial effect was due to the overproduction of ROS, triggered by UV-A irradiation. As a potential antimicrobial agent, polymeric [Ru(CO)$_2$Cl$_2$]$_n$ offers key advantages, including ease of synthesis, high solubility, and low cytotoxicity. Furthermore, the antibacterial efficiency of [Ru(CO)$_2$Cl$_2$]$_n$ is not paralleled in other ruthenium complexes whose activity has been previously reported. Thus, the present investigation opens the possibility of exploring polymeric ruthenium precursors for bacterial infection treatments with minimal cytotoxicity. These compounds may represent an important advance in the development of new antimicrobials without using expensive hydrophobic ligands due to the possibilities for chemical modification, high purity and low-cost in comparison to ruthenium complexes with organic ligands.

**Acknowledgements.** This project has received funding from the European Union's Horizon 2020 research and innovation programme under the Marie Skłodowska-Curie grant agreement No 754382, GOT ENERGY TALENT, the Spanish Government (CTM2016-74927-C2-1/2-R) and the Universidad de Alcala (CCG19/CC-037). The content of this article does not reflect the official opinion of the European Union. Responsibility for the information and views expressed herein lies entirely with the authors.

# Supplementary Material

# Polymeric ruthenium precursor as a photoactivated antimicrobial agent

Srabanti Ghosh[1,*], Georgiana Amariei[2], Marta E. G. Mosquera[1,*], Roberto Rosal[2]

[1] Department of Organic and Inorganic Chemistry, Instituto de Investigación en Química "Andrés M. del Río" (IQAR), Universidad de Alcalá, Campus Universitario, 28805, Alcalá de Henares, Madrid-Spain
[2] Department of Chemical Engineering, Universidad de Alcalá, Campus Universitario, 28805, Alcalá de Henares, Madrid-Spain

* Corresponding authors: Dr. Marta E. G. Mosquera, martaeg.mosquera@uah.es & Dr. Srabanti Ghosh, srabanti.ghosh@uah.es

## Materials and methods (supplementary details)

### Manipulation of microorganisms

The microorganisms were preserved at -80 °C in glycerol (20% v/v) until use. The microorganisms were reactivated in nutrient broth (NB, 10 g L$^{-1}$ peptone, 5 g L$^{-1}$ sodium chloride, 5 g L$^{-1}$ meat extract and, for solid media, 15 g L$^{-1}$ powder agar, pH 7.0 ±0.1) by incubation at 37°C under shaking at 100 rpm. Inoculums were prepared in 1/500 NB with $10^8$ cells mL$^{-1}$ for *E. coli* and $10^{10}$ cells mL$^{-1}$ for *S. aureus* (followed by optical density at 600 nm) ensuring the exponential growth phase of both microorganisms during contact experiments. The antimicrobial and antifouling effects of material were evaluated according to the standardized ISO 22196 test, followed with minor modifications.

### Irradiances

Selected irradiances were based on NASA Surface Meteorology and Solar Energy Database (https://power.larc.nasa.gov/). The average daily incident insolation reported for a horizontal surface at the latitude of Madrid was 6.1 kW-h m$^{-2}$ in Spring-Summer and 2.7 kW-h m$^{-2}$ in Winter-Fall, from which 5.2 % and 5.7 % respectively correspond to UV irradiation. As a conservative assumption, the irradiation time of the LED lamp was adjusted to irradiate with one third of the value indicated by the solar irradiation database for summer-spring (2.0 kW-h m$^{-2}$) or winter-fall (1.0 kW-h m$^{-2}$).

### Statistics

A one-way ANOVA coupled with Tukey's HSD (honestly significant difference) post-hoc test was performed for comparison of means. Statistically significant differences were considered to exist when p-value < 0.05.